\documentclass[sigconf]{acmart}

\renewcommand\footnotetextcopyrightpermission[1]{}

\usepackage{booktabs} 
\usepackage{balance}

\setcopyright{rightsretained}

\acmConference[SysML'18]{SysML Conference}{February 2018}{Stanford, CA, USA}
\acmYear{2018}
\copyrightyear{2018}

\begin{document}
\title{Representation Learning for Resource Usage Prediction}

\author{Florian Schmidt}
\orcid{0000-0003-2307-2579}
\affiliation{\institution{NEC Laboratories Europe}}
\email{florian.schmidt@neclab.eu}

\author{Mathias Niepert}
\affiliation{\institution{NEC Laboratories Europe}}
\email{mathias.niepert@neclab.eu}

\author{Felipe Huici}
\affiliation{\institution{NEC Laboratories Europe}}
\email{felipe.huici@neclab.eu}

\maketitle

\section{Introduction}

Creating a model of a computer system that can be used for tasks such as predicting future resource usage and detecting anomalies is a challenging problem.
Most current systems rely on heuristics and overly simplistic assumptions about the workloads and system statistics.
These heuristics are typically a one-size-fits-all solution so as to be applicable in a wide range of applications and systems environments.
This limitation is all the more striking considering the wide range of problems that could be approached with a more sophisticated model of a computing system:
for example, resource allocators, both process schedulers in OSs and orchestrators in clusters\cite{linux-cfs,fair-scheduling,quincy-sosp,xen-schedulers}, use simple heuristics, and still often struggle to get performance right\cite{wasted-cores,xen-credit-scheduler-comparison};
and monitoring systems whose objective is the detection of anomalies have used some machine-learning approaches in network-based scenarios\cite{hybridml-network-anomaly,sommer:outside,buczak:mlidsurvey}, but much less so in the more systems-heavy domain.
Tailoring the prediction models to specific situations, however, can be extremely complex: they have to take into account the interplay of systems components and concurrently running heterogeneous applications, while being able to adapt to a dynamically and often abruptly changing state of the system.
Considering developing generic heuristics is already an extremely time-consuming task, creating tailor-made solutions by hand is rarely worth the effort.

However, there are several recent developments that bring us closer to developing systems models that could perform much better than existing generic methods based on simple heuristics. Machine learning is becoming more effective and efficient at learning from large amounts of data. Moreover, we have a much better understanding of ways to embed heterogeneous feature types (categorical, numerical, structured, temporal) into a joint representation amenable to downstream tasks. If we extract and collect the right input data, we may be able to automatically create tailor-made models that outperform generic heuristics.

With this paper, we present our ongoing work of integrating systems telemetry ranging from standard resource usage statistics to kernel and library calls of applications into a machine learning model. Intuitively, such a ML model approximates, at any point in time, the state of a system and allows us to solve tasks such as resource usage prediction and anomaly detection. To achieve this goal, we leverage readily-available information that does not require any changes to the applications run on the system.
We train recurrent neural networks such as Long Short-Term Memory (LSTM) neural networks~\cite{Hochreiter:1997} to learn a model of the system under consideration. As a proof of concept, we train models specifically to predict future resource usage of running applications.

\begin{figure*}
	\includegraphics[width=0.82\textwidth]{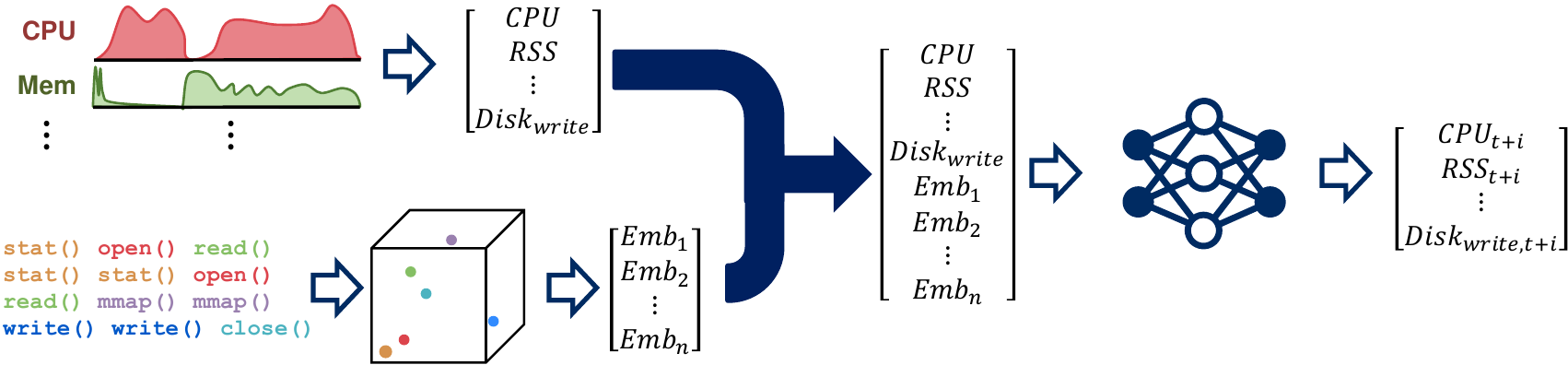}
	\vspace{-6pt}
	\caption{To predict resource usage, we collect typical telemetry data (top left), as well as the application's system calls (bottom left), over a time period $t$. The variable number of calls is transformed into a fixed-size vector via a word embedding. The two vectors are combined and used as input for an LSTM that then predicts the resource usage at some future time period $t+i$.}
	\vspace{-3pt}
	\label{fig:chronos-arch}
\end{figure*}

\section{Data Collection}

To learn a model of a system that is good at predicting future resource usage, we need to collect data about the present.
One obvious approach is to collect data about the resources that we want to predict such as CPU and memory usage statistics.
This follows the idea that, in many cases, the previous values of resource usage will have at least some influence on current resource usage.
For example, memory consumption will often increase or decrease gradually over time.
CPU usage is often more spiky, but even here, for many processes, phases of low activity and high activity will be apparent.
Such resource usage information is easily available on Unix-like systems; however, it is generally accounted for on a per-process basis.
This is useful for process scheduling but for more coarse-grained scheduling of jobs or services, which comprise several processes run in sequence or in parallel, we will need aggregate measurements.
For this, we monitor the process group, that is, all processes spawned by one initial process (that do not specifically request to leave the group).
Aggregation is slightly cumbersome because there is no easy way to look up all processes belonging to a group given the group ID; it instead requires traversing all processes and asking them for the process group they belong to.
This, and the resource information, can be collected from by reading it from \texttt{/proc/<pid>/stat}.
Alternatively, a new \texttt{cgroup} can be created and the initial process spawned into it.
The resource requirements of the process group are then the resources used by the \texttt{cgroup}.

However, these high-level usage statistics alone provide no deeper insights into the state of a process.
It would be useful to have at least a rough understanding of \textit{what} a process ``is doing" at runtime.
Unfortunately, the possibilities here are limited if we want to stay generic and not require ancillary or internal information that is specific to a certain problem domain (such as information about input data), or requiring specific compiling or linking steps.
Using a profiler to measure which functions are being run for how long, for instance, requires a symbol table which is not always available (stripped).
There are options, however, to inspect program behavior without requiring such additional information.
By analyzing the \texttt{system calls} that a program performs, we can get a rough understanding of what a process is doing and this information is always available, because it does not rely on code annotation or additional symbols.
Some system calls also have an obvious relationship with certain kinds of resources.
For example, the \texttt{write}, \texttt{read} and similar system calls work on files or sockets, which translates into disk or network I/O.
If we want to predict I/O, the relationship is obvious; but even for CPU usage, there is a strong relationship: for example, disk I/O often correlates with low CPU usage, since the process is waiting on I/O accesses to the finished.

System calls are also easily traced: \texttt{strace}\cite{strace} is a standard tool available on Unix-like systems, and these days, its overhead is low -- a few percent when a moderately high number of system calls occurs, to virtually none when there are no system calls happening.
In case of workloads with extremely high numbers of system calls, \texttt{perf}\cite{perf} can be used to sample system calls instead of tracing every single one, further reducing the overhead.
Conversely, if a higher level of detail is required, \texttt{ltrace}\cite{ltrace} can be used instead to catch all library interactions.
As a proof of concept, we developed a ML model that integrates usage statistics and sequences of system calls into a joint representations.

\section{Data Preprocessing}

In order to prepare the data such as usage statistics as input for the ML models, we need to discretize it into time intervals for several reasons.
First, some data only makes sense as values over a time period: what was the CPU utilization in the last second? How many bytes were written to disk?
Second, for the eventual goal of resource allocation, we will also have to predict resource usage over a a time period that the scheduler uses as time slice.
Finally, calculating each resource usage over a certain time period provides us a fixed-size value: each information can be interpreted as a single value which then can all be combined into an input vector of fixed size.
For system calls, however, such a fixed-size representation is not straightforward to generate.
System calls occur at (seemingly) random times and are discrete events as opposed to continuous numerical values.
Within a time period of a second thousands of system calls, or none, can occur.
Fortunately, to transform sequences of system calls into a fixed-sized vector representation, we can use representation learning approaches for sequence data such as the word2vec skip-gram model~\cite{Mikolov:2013}.
Instead of applying these representation learning approaches to sequences of words to learn meaningful vector representations, we can apply these methods to sequences of system calls (and also other types of event sequences occurring a system) to learn representations of these events.
To collect a corpus of system call sequences, we ran \texttt{strace}\cite{strace} on the various types of applications run on the system under consideration.
We then used that data to learn system call embeddings through a model similar to the skip-gram model~\cite{Mikolov:2013}.
As a result, we can take all system calls occurring within a time period, interpret them as a ``sentence," and use the event embeddings to create a fixed-size vector representation. Since we now have a fixed-sized representation of the resource usage statistics and a fixed-size representation for system calls, we can directly use this data to train a ML systems model, as shown in Figure~\ref{fig:chronos-arch} that depicts the overall architecture.

\begin{figure}
\includegraphics[width=\columnwidth]{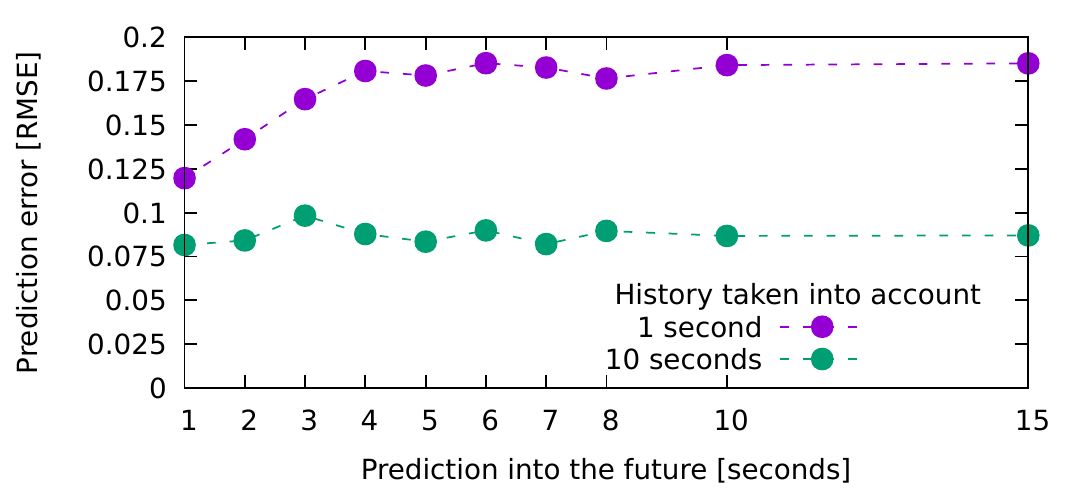}
\vspace{-12pt}
\caption{Looking farther into the future tends to increase the prediction error, but taking more history into account mitigates this effect.}
\vspace{-9pt}
\label{fig:results}
\end{figure}

\section{Neural Networks for Systems Modeling}

The objective of this work is to learn and maintain a model of a computing system on a particular level of abstraction. In the end, all systems are state-based and, given a current state, we want to use the model of the system to make predictions about its and its applications future behavior. Several recent neural network based machine learning architectures maintain some sort of internal state. Examples are recurrent networks such as LSTMs~\cite{Hochreiter:1997} and variants~\cite{pmlr-v48-danihelka16}, memory networks~\cite{Weston:2014,Sukhbaatar:2015}, and neural Turing machines~\cite{Graves:2014}, to name but a few.
We are taking advantage of these methods by developing a system model that maintains a vector (hidden) representation of the current state of the system and is trained so as to minimize the expected error (here: the root-mean-square error (RMSE)) of predicting future resource usage. To keep the model simple and for the use case of resource usage prediction, we train an LSTM with the collected and preprocessed data consisting of past usage statistics and system calls. The learning of system calls embeddings can be performed as a preprocessing step or within an end-to-end architecture.

We conducted some preliminary experiments by collecting system calls from various applications to create the system call corpus.
We then collected the resource usage and system calls of a scientific computing toolchain that executed a number of bash and python scripts, which interleaved I/O- and CPU-heavy phases.
Finally, we embedded the system calls and trained an LSTM with the data.
The model is trained to minimize the RMSE of the CPU usage (as a value between 0 and 1) $i$ seconds into the future. Figure~\ref{fig:results} shows the results, varying both how far to predict into the future, and how much history to take into account for the prediction.

\textbf{Acknowledgments}---This project has received funding from the European Union's Horizon 2020 research and innovation programme under grant agreement No 761592.

\bibliographystyle{ACM-Reference-Format}
\bibliography{references}

\end{document}